\documentclass[aps,twocolumn,fleqn,superscriptaddress]{revtex4}

\usepackage{graphicx,color}
\usepackage{amsmath,amssymb,amsfonts}

\newcommand{\be}{\begin{equation}}
\newcommand{\ee}{\end{equation}}

\newcommand{\bea}{\begin{eqnarray}}
\newcommand{\eea}{\end{eqnarray}}
\newcommand{\eq}[1]{(\ref{#1})}

\newcommand{\half}{{1\over2}}

\def\be{\begin{equation}}
\def\ee{\end{equation}}
\def\beq{\begin{eqnarray}}
\def\eeq{\end{eqnarray}}
\def\a{\alpha}

\def\vp{\varphi}

\def\({\left (}
\def\){\right )}
\def\[{\left [}
\def\[{\right ]}

\def\a{\alpha}

\input amssym.def
\input amssym.tex

\def\del{\partial}
\def\eps{\epsilon}

\newcommand{\arXiv}[1]{\href{http://www.arXiv.org/abs/#1}{#1}}
\usepackage[colorlinks=true, linkcolor=blue, bookmarks=true]{hyperref}

\begin{document}

\title{ Inhomogeneous Thermalization in Strongly Coupled Field Theories}

\author{V.~Balasubramanian}
\affiliation{David Rittenhouse Laboratory, University of Pennsylvania, Philadelphia, PA 19104, USA}
\affiliation{Laboratoire de Physique Th\'{e}orique, \'{E}cole Normale Sup\'{e}rieure, 75005 Paris, France} 
\author{A.~Bernamonti}
\affiliation{Instituut voor Theoretische Fysica, KU Leuven,
Celestijnenlaan 200D, B-3001 Leuven, Belgium}
\author{J.~de~Boer}
\affiliation{Institute for Theoretical Physics, University of Amsterdam,
1090 GL Amsterdam, The Netherlands }
\author{B.~Craps}
\affiliation{Theoretische Natuurkunde, Vrije Universiteit Brussel, and
Int.~Solvay Inst., B-1050 Brussels, Belgium}
\author{L.~Franti}
\affiliation{Helsinki Institute of Physics \& Dept.~of Physics,
FIN-00014 University of Helsinki, Finland}
\author{F.~Galli}
\affiliation{Theoretische Natuurkunde, Vrije Universiteit Brussel, and
Int.~Solvay Inst., B-1050 Brussels, Belgium}
\author{E.~Keski-Vakkuri}
\affiliation{Helsinki Institute of Physics \& Dept.~of Physics,
FIN-00014 University of Helsinki, Finland}
\affiliation{Department of Physics and Astronomy, Uppsala University, SE-75108 Uppsala, Sweden}
\author{B.~M\"uller}
\affiliation{Department of Physics, Duke University, Durham, NC 27708, USA}
\affiliation{Brookhaven National Laboratory, Upton, NY 11973, USA}
\author{A.~Sch\"afer}
\affiliation{Institut f\"{u}r Theoretische Physik, Universit\"{a}t Regensburg, D-93040 Regensburg, Germany}

\begin{abstract}
To describe theoretically the creation and evolution of the quark-gluon plasma, one typically employs three ingredients: a model for the initial state, non-hydrodynamic early time evolution, and hydrodynamics. In this paper we study the non-hydrodynamic early time evolution using the AdS/CFT correspondence in the presence of inhomogeneities. We find that the AdS description of the early time evolution is well-matched by free streaming. Near the end of the early time interval where our analytic computations are reliable, the stress tensor agrees with the second order hydrodynamic stress tensor computed from the local energy density and fluid velocity.
Our techniques may also be useful for the study of far-from-equilibrium strongly coupled systems in other areas of physics.
\end{abstract}

\maketitle

Strongly coupled quantum liquids are studied in many experimental settings: (a) quark-gluon matter created in ultrarelativistic heavy ion collisions, (b) strongly correlated electrons in metals, cuprates and heavy-fermion materials, and (c) condensates of ultra-cold atoms.  The holographic AdS/CFT or gauge-gravity framework has provided new insights into the dynamics of such fluids.  One development is the fluid/gravity correspondence \cite{Hubeny:2011hd} where one can derive hydrodynamic equations from Einstein's equations in a particular long wavelength limit, and even get  complete second order hydrodynamic equations \cite{Baier:2007ix,Bhattacharyya:2008jc} for conformal relativistic fluids. Another interesting question is how and when a far-from-equilibrium initial state approaches a regime where hydrodynamics becomes a good approximation. This article studies this question in an analytically tractable model that exhibits inhomogeneities. 

Experiments at RHIC and at the LHC have shown that already at very early times, at most 1 fm/c, the matter produced in heavy ion collisions shows collective behavior in agreement with viscous hydrodynamics.  The argument about the presence of hydrodynamic behavior of the quark-gluon plasma created in heavy ion collisions rests primarily on two observations. First, one observes a $\cos(2\phi)$ correlation between the azimuthal momentum direction of produced hadrons and the collision plane, which is known as ``elliptic'' flow, see \cite{Adare:2006ti,Adamczyk:2013gw,ALICE:2011ab,ATLAS:2012at} and references therein. This phenomenon can also be deduced from the azimuthal two-particle correlations among emitted hadrons, and is sometimes referred to as the ``ridge'' or ``double-ridge'' effect in studies in Pb+Pb collisions at the LHC \cite{ATLAS:2012at,Chatrchyan:2012wg,Abelev:2009af,Aamodt:2011by}. 

The second observation is related to event-by-event fluctuations \cite{Heinz:2013th,Adare:2012kf}. Experimentally it was found for central heavy ion collisions that odd Fourier coefficients of the flow are not much smaller than even ones \cite{ATLAS:2012at,Abelev:2012di,Chatrchyan:2012wg}. 
By a symmetry argument, odd coefficients can only be generated by fluctuations, so one is forced to conclude that fluctuations have large amplitudes. 
The parton saturation model for the initial nuclear state suggests also that the fluctuations are of short range in the plane transverse to the beam axis (of order of the inverse saturation scale $1/Q_s \sim$ 0.2 fm  \cite{Muller:2011bb}).  Detailed simulations \cite{Schenke:2012hg,Gale:2012rq,Bzdak:2013zma}  use phenomenological models (such as the Color Glass Condensate model) to compute the initial energy deposition and  to evolve it for a short amount of time (of order 0.4 fm/c) using the classical Yang-Mills equations.  These studies give excellent fits to all flow coefficients by assuming that large-amplitude, small-range fluctuations subsequently evolve hydrodynamically. Note that for times short compared to the scale set by the background Yang-Mills field, typically of order 0.2 fm/c, classical Yang-Mills theory is kinetic energy dominated and should therefore be well-approximated by a free-streaming model. Alternate approaches start from fluctuations in nucleon positions and the energy deposition in individual nucleon-nucleon collisions \cite{Alver:2010gr,Petersen:2010cw} and use either free-streaming of particles \cite{Qin:2010pf} or hydrodynamics of  anisotropic fluids \cite{Martinez:2010sc,Ryblewski:2012rr} to bridge the gap to the onset of viscous hydrodynamics.  An important open question, which motivated the present work, is to what extent this ``gluing'' of a phenomenological model describing the initial evolution to viscous hydrodynamics can be justified.

\begin{figure}[ht]
\includegraphics[width= 0.28\textwidth]{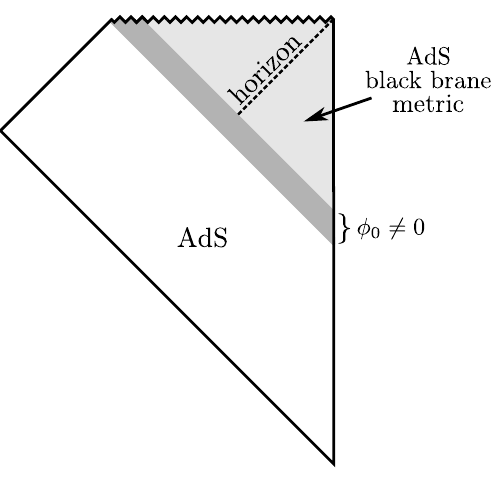}
\caption{\label{fig:collapse}A schematic representation of the dynamical collapse process first studied in \cite{Bhattacharyya:2009uu}.}
\end{figure}

Thermalization in strongly coupled conformal field theories with a gravity dual corresponds to black brane formation in asymptotically anti-de Sitter (AdS) spacetimes \footnote{By black brane formation we mean the formation of a static black brane. This is generically preceded by the formation of a far-from-equilibrium black brane state, which subsequently relaxes to a static black brane.}. To study gravitational collapse, one generically needs numerical general relativity, but interesting situations exist in which analytical calculations are possible. Consider a massless scalar field minimally coupled to gravity in $d+1$ space-time dimensions with negative cosmological constant. In this setting, the authors of \cite{Bhattacharyya:2009uu} considered the effect of a homogeneous boundary source on the bulk geometry of an asymptotically AdS space-time. Specifically, they briefly $(0<t<\delta t)$ turned on a homogeneous source $\phi_0(t)=\epsilon\tilde \phi_0(t)$ for a marginal boundary operator corresponding to the massless scalar field in the bulk and solved the field equations perturbatively in $\epsilon$. In $d+1$ bulk dimensions and to second order in  $\epsilon$, they found the AdS-Vaidya metric
\be\label{AdSVaidya}
ds^2 =  - \(r^2 - \frac{M(v)}{ r^{d-2}} \) dv^2+ 2 dv dr + r^2 \, d\vec x^2 +{\cal O}(\epsilon^4)\, ,
\ee
where $M(v)$ vanishes for $v<0$ and is otherwise of order $\epsilon^2/(\delta t)^d$ (we have set $R_{\text {AdS}} =1$). For odd $d$, $M$ is constant for $v>\delta t$ and the geometry then describes a black brane with temperature of order $\epsilon^{2/d}/\delta t$. This space-time describes a shell of null dust falling in from the boundary of AdS and collapsing into a black brane. At the boundary, the Eddington-Finkelstein coordinate $v$ coincides with the field theory time $t$. A schematic representation of the process is depicted in Fig.~\ref{fig:collapse}.  
The weak-field approximation amounts to considering injection times short compared to the inverse temperature of the black brane to be formed.

Naive perturbation theory in $\epsilon$ is reliable for $v\ll 1/T$, but diverges for $v\gg 1/T$. A resummed perturbation theory can be defined by expanding around the AdS-Vaidya geometry (\ref{AdSVaidya}) rather than AdS, i.e., by working exactly in $M(v)$ and perturbatively in all other occurrences of $\epsilon$  \cite{Bhattacharyya:2009uu}. (This is analogous to absorbing temperature-dependent masses in propagators in thermal perturbation theory.) This approach is technically more complicated, but it was shown in \cite{Bhattacharyya:2009uu}  that resummed perturbation theory is reliable everywhere outside the event horizon. For late times $(v\gg 1/T)$, observables decay exponentially to their thermal values, with relaxation times of order of the inverse temperature. 

The AdS-Vaidya model has been very useful as a simple, tractable model of holographic thermalization. In \cite{AbajoArrastia:2010yt, Albash:2010mv, Balasubramanian:2010ce, Balasubramanian:2011ur}, it was found that in this model thermalization proceeds at the speed of light, at least for the range of length scales studied in \cite{ Balasubramanian:2010ce, Balasubramanian:2011ur}. (For $d>2$, it was found in \cite{Liu:2013iza} that on length scales larger than the inverse temperature, thermalization happens at a smaller speed.) However, the initial state of heavy ion collisions is quite anisotropic and inhomogeneous. It is characterized by a strong asymmetry between longitudinal and transverse pressure (the former may even be negative initially due to the presence of strong longitudinal gauge fields) and by large density fluctuations in the transverse direction. The effect of both can be studied with a refined AdS/CFT treatment.  The effect of the pressure anisotropy was studied under the assumption of longitudinal boost invariance and transverse homogeneity in \cite{Chesler:2009cy,Beuf:2009cx, Heller:2011ju}, where it was found that hydrodynamic behavior is reached on time scales of order $0.3-0.5$ fm/c for many different initial conditions.  This ``hydroization'' is not equivalent to complete thermalization because viscous hydrodynamic behavior at early times in a boost invariant expansion implies a rather large pressure anisotropy and thus strong deviation from local thermal equilibrium. In the present work, we analyze the second aspect, namely the question how the approach to hydrodynamic behavior is affected by local density fluctuations, which has not been studied in detail so far.  

In a companion paper~\cite{Balasubramanian:2013oga}, we generalize the construction of \cite{Bhattacharyya:2009uu} to the case of an inhomogeneous scalar field source at the boundary (with the solution coinciding with pure AdS before the source is turned on). As in \cite{Bhattacharyya:2009uu}, we solve the equations of motion in a perturbative expansion in the amplitude of the scalar field boundary value, which is here allowed to depend on the spatial coordinates. In order to retain analytical control, we make the simplifying assumption (suggested in \cite{Bhattacharyya:2009uu}) that the scale of spatial variations is large compared with all other scales. (In a heavy ion context, this assumption is not justified for fluctuations at a scale $1/Q_s$, which is smaller than the inverse temperature, so some extrapolation will be needed to make contact with those.) As in \cite{Bhattacharyya:2009uu}, the case of a four-dimensional bulk space-time turns out to be technically simpler than the five-dimensional one. Since in heavy ion collisions azimuthal anisotropies are studied in the directions transverse to the beam, a four-dimensional bulk geometry may not constitute a serious limitation. More realistic models would involve nearly boost invariant geometries or colliding inhomogeneous shock waves, but for computational tractability these will not be studied in the present paper. As a further technical simplification, we consider the simplest case involving spatial dependence on a single coordinate, in addition to the radial coordinate in the bulk. All in all, we consider an asymptotically AdS$_{4}$ geometry with inhomogeneities along a single spatial direction.

Specifically, we construct the bulk solution for the scalar field and the metric up to second order in the amplitude of the source that drives the gravitational collapse and up to fourth order in the gradient expansion. For instance, up to second order in the amplitude of the source $\varphi(v,x)$ and up to fourth order in spatial gradients, solving the coupled equations of motion for the metric and  scalar field $\phi$ gives:
\be
\phi=\varphi(v,x)+\frac{\dot\varphi(v,x)}{r}+\frac{\varphi^{\prime\prime}(v,x)}{2r^2}+\frac{\int_{-\infty}^v d\tau\varphi^{\prime\prime\prime\prime}(\tau,x)}{8r^3}+\ldots
\ee
Similar expressions can be obtained for the various metric components.
Our solutions are reliable for times $v$ short compared to the inverse local temperature $T(x)$ (of the black brane to be formed) and local scales of variation $L(x)$ large compared to the inverse temperature, $v\ll 1/T(x) \ll L(x)$.  From these bulk solutions, we extract the expectation value of the boundary energy-momentum tensor, and study its time-evolution after the inhomogeneous energy injection. Specifically, $\langle T_{\alpha\beta}\rangle\sim g_{(3)\alpha\beta}$, where $ g_{(3)}$ is the $1/r$ coefficient of the bulk metric in Fefferman-Graham coordinates. For instance, up to second order in the amplitude expansion and in the derivative expansion,
\bea \label{tttxt}
&&T_{tt}(t,x) =- \frac{1}{16 \pi G_{N}} \left\{  \frac{1}{2} \int_{-\infty}^{t} d\tau~ \Big[ 2\dot\vp(\tau,x)\dddot \vp(\tau,x)\right.\nonumber\\
&&  -\( \dot\vp'(\tau,x)\)^2   -  4 \dot\vp(\tau,x)\dot\vp''(\tau,x) + 2 \ddot\vp(\tau,x) \vp''(\tau,x)   \nonumber \\
&&   + 2\ddot\vp'(\tau,x)\vp'(\tau,x)+ 2\frac{\del}{\del{x}} \int_{- \infty}^{\tau} ds  [ \dot\vp(s,x)\ddot\vp'(s, x)   \nonumber\\
&& \left.  -    \int_{- \infty}^{s} dw~\ddot\vp'(w,x)\ddot\vp(w, x) ]  \Big] \right\}  \, .
\eea
To compute the evolution for later times, a resummation of perturbation theory would be required, but this is beyond the scope of the present work, in which we restrict ourselves to the early-time evolution. From the energy-momentum tensor, we compute the local velocity $V(t,x)$ (i.e., the velocity relative to the lab frame of the local rest frame in which the stress-tensor becomes diagonal), as well as the energy density $\varepsilon(t,x)$ and pressure components $p_x(t,x),\, p_y(t,x)$ in the local rest frame.

\begin{figure}[h]
\includegraphics[width = 0.4 \textwidth]{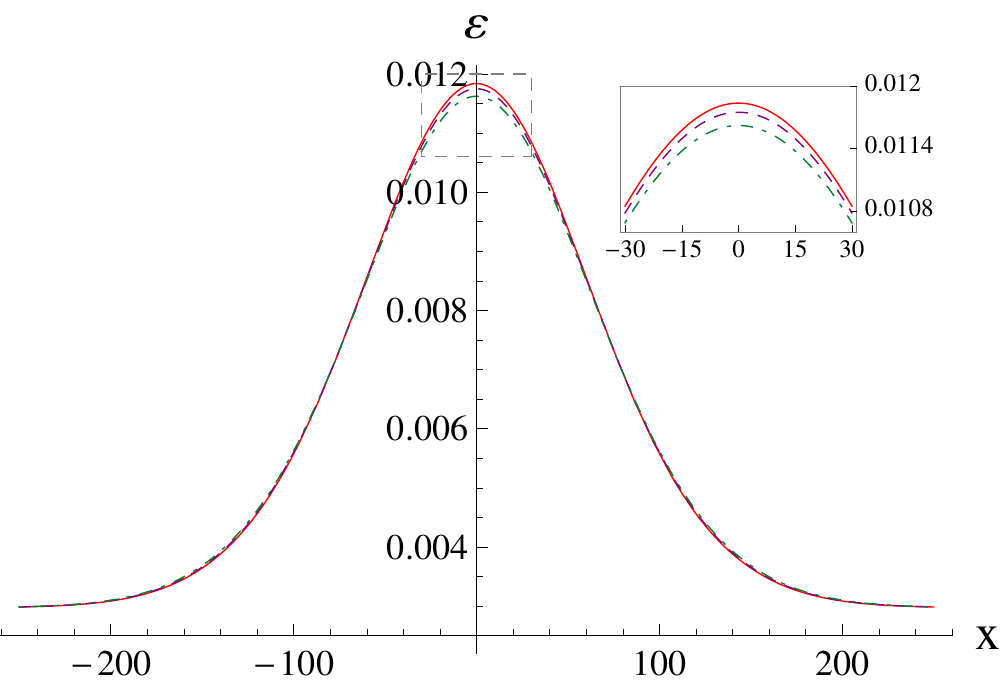} 
\includegraphics[width = 0.4 \textwidth]{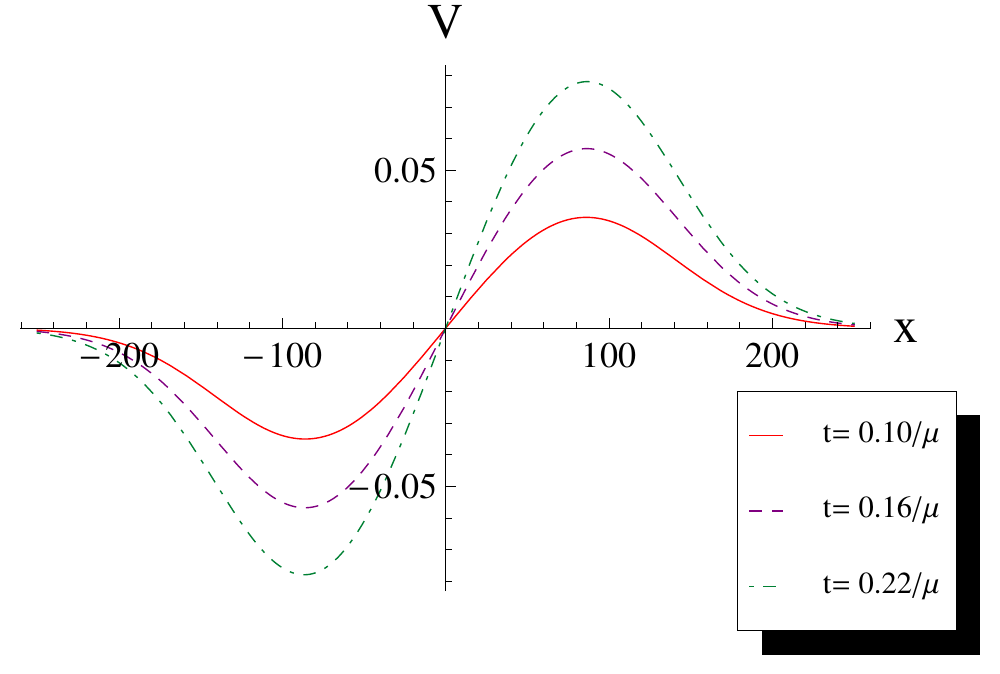}
\caption{\label{fig:palpha}The energy density in the local rest frame $\varepsilon =p_x +p_y$ (rescaled by $16\pi G_{N}$)  and the plasma  local velocity $V$ for the solution with  $\nu =0.5$, $\sigma^2 = 0.1$,  $\mu = 0.01$ and $\eps  = 0.005$. As  time increases,  $\varepsilon$ decreases in the region around  $x=0$ and increase for large $|x|$, while the velocity curves depart from the $x$-axis.
}
\end{figure}

In order to study the energy-momentum tensor in more detail, we use a Gaussian profile in the laboratory frame to mimic the time dependence of the source, $\varphi(t, x) =   \eps  \, u(x) \exp[-(t -\nu)^2/\sigma^2]$:  when  $t -\nu$ is larger than a  few $\sigma$, a time which  can be  identified with the injection time  $\delta t$, the scalar source  can be considered to vanish for all practical purposes. We consider the simple case where the spatial profile $u(x)$ has the form of a Gaussian with a homogeneous background, $u(x) = 1+   \exp(- \mu^2 x^2)$. (More general combinations of Gaussians and background are studied in \cite{Balasubramanian:2013oga}.)
Our results for the energy density in the local rest frame and for the local velocity are shown in Fig.~\ref{fig:palpha}. 
In order to gain some insight into these results, we next compare them with the evolution of the energy-momentum tensor in a free-streaming model as well as in hydrodynamics.  

\begin{figure}
\includegraphics[width=0.85\linewidth]{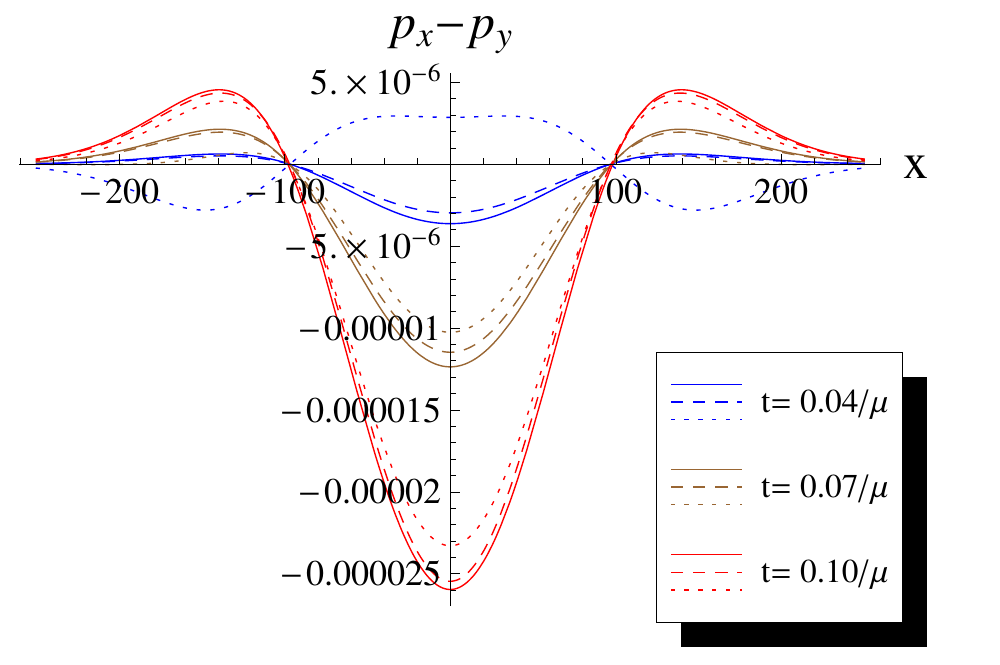}
\caption{\label{fig:anisotropies}Comparison of the pressure anisotropies (rescaled by $16\pi G_{N}$): AdS/CFT result in solid line, free streaming curves in dashed, second order hydrodynamics in dotted lines. The times plotted are shorter than those in Fig.~\ref{fig:palpha} to ensure that the corrections to $p_x$ and $p_y$ coming from higher orders in the derivative expansion are negligible compared to the pressure anisotropies.
}
\end{figure}

Our free streaming model is obtained by assuming that the distribution is massless noninteracting dust, composed of particles moving at the speed of light. In terms of the phase space distribution $f(t, \vec x, \vec k)$, the  components of the stress-energy tensor are given by
\be
T^{\alpha\beta}(t, \vec x) = \int d^2k \frac{k^\alpha k^\beta}{k^0} f(t, \vec x, \vec k) \,.
\label{eq:Tmunu-streaming}
\ee
If we assume the phase space distribution at $t=0$ to have the factorized form $f(\vec x, \vec k) = n(\vec x) F(\vec k) = n(x) F(k)$, then at a later time 
\be
f(t, \vec x, \vec k) = n(\vec x-\vec v t) F(\vec k) = n(x-v_x t) F(k),
\ee   
where $\vec v=\vec k/k^0$ is the particle velocity, and $F(k)$ only appears in \eq{eq:Tmunu-streaming} in an overall normalization factor. We choose $n(x)$ such that just after the energy injection has ended, the energy density in the laboratory frame coincides with the energy density we computed using AdS/CFT, and we compare the pressure anisotropies in the local rest frame in Fig.~\ref{fig:anisotropies}. 

Given the local velocity $V(t,x)$ and the energy density in the local rest frame $\varepsilon(t,x)$, we can ask what the stress-tensor would be if (first or second order) hydrodynamics were valid at a given time $t$. Agreement with the stress tensor we computed using AdS/CFT would be a necessary condition for the validity of a (first or second order) hydrodynamical description from time $t$ onwards.
First order hydrodynamics requires the equation of state and the shear and bulk viscosities, which were obtained in \cite{Hubeny:2011hd,Bhattacharyya:2008mz} for the three-dimensional CFT under consideration. The first-order hydrodynamical stress tensor reads
\be \label{eq:Tvisc}
T_{\rm viscous}^{\a\beta} =  \( \varepsilon + p_{\rm ideal } \) u^\a u^\beta + p_{\rm ideal } \, \eta^{\a\beta} + \Pi^{\a\beta}_{(1)}\, .
\ee
Here $p_{\rm ideal }=\varepsilon/2$ and $u^\alpha$ is the local three-velocity of the fluid (which we determine from $V(t,x)$). The first order viscous contributions to the energy-momentum tensor in flat three-dimensional space are given by \cite{Hubeny:2011hd}
\be 
\Pi^{\a\beta}_{(1)} \equiv - 2 \eta \sigma^{\a \beta} - \zeta \theta P^{\a\beta}\,,
\ee
where $P^{\a \beta} \equiv u^\a u^\beta + \eta^{\a\beta}$ is the projector onto space in the local fluid rest frame,
\be
\sigma^{\a\beta} \equiv P^{\a\rho} P^{\beta\sigma} \left(\del_{(\rho}u_{\sigma)} - \frac 1 2 P_{\rho\sigma} \theta \right)
\ee
is the fluid shear tensor and $\theta \equiv \del_\rho u^\rho$ is the expansion. The shear viscosity $\eta$ and bulk viscosity $\zeta$ are \cite{Hubeny:2011hd,Bhattacharyya:2008mz}
\be \label{eq:minimalshear}
\eta = \frac{1}{16 \pi G_N} \left(\frac{4}{3} \pi T\right)^2\,, \qquad \zeta = 0\, ,
\ee
where we define a local ``temperature'' $T(t,x)$ by using the thermodynamic relation that would be valid in equilibrium in the local rest frame,
\be
\varepsilon=  \frac{2}{16\pi G_N} \(\frac{4}{3} \pi T\)^3\, .
\ee
Using these ingredients, we have computed in \cite{Balasubramanian:2013oga} the pressure anisotropy in first order hydrodynamics. While qualitatively in agreement with our AdS/CFT results, it is too large to be in good quantitative agreement, so we conclude that first order hydrodynamics is not applicable in the early time window we can study \cite{Balasubramanian:2013oga}.

Second order hydrodynamics for relativistic conformal fluids was derived in \cite{Baier:2007ix, Bhattacharyya:2008jc}, providing a nonlinear generalization of M\"uller-Israel-Stewart theory. The second order contribution to the dissipative part of the stress tensor is 
\be\label{second}
\Pi^{\alpha\beta}_{(2)} = 
\eta \tau_{\Pi} \left[  \mbox{}^{\langle} D\sigma^{\alpha\beta\rangle} + \half \sigma^{\alpha\beta} \theta \right] + \cdots
\ee
where $\langle \rangle$ denotes the transverse traceless part  \cite{Baier:2007ix}, and  $``+\cdots"$ refers to terms that are not relevant here, as they either contain the Riemann/Ricci tensor or terms that vanish for irrotational flow. In (\ref{second}) $D$ is the directional derivative along the 3-velocity, $D \equiv u^\mu \partial_\mu = -u_t\partial_t+u_x\partial_x$, and the new parameter $\tau_{\Pi}$ is the relaxation time for shear stress. For a strongly coupled conformal fluid it depends on harmonic numbers \cite{Hubeny:2010ry}, in 2+1 dimensions 
\be\label{relaxtime}
\tau_{\Pi} = \frac{3}{4\pi T}\left[ 1 + \frac{1}{3} {\rm Harmonic}\left(-\frac{1}{3}\right)\right]  .
\ee
A short calculation yields the hydrodynamic pressure anisotropy
\be\label{delta-pH}
p_x^{\rm (H)} - p_y^{\rm (H)} = 2 \eta \left[- \theta +  \tau_{\Pi} \( D\theta + \frac 1 2 \theta^2\) \right]\,.
\ee
The pressure anisotropies obtained from second order hydrodynamics are compared to those from AdS/CFT and from free-streaming in Fig.~\ref{fig:anisotropies}.
Throughout the time period in which we can trust our holographic solution free-streaming provides an excellent description of our holographic results. At the end of this time period, there is also consistency with second order viscous hydrodynamics. We are not able to make predictions for later times and cannot exclude the possibility that the holographic solution might again deviate from that of local viscous hydrodynamics at some later time. However, our results are fully compatible with the assumption that initial free streaming behavior turns into local hydrodynamic one within times smaller than $1/T$. 

The reader may wonder why free streaming can reproduce the behavior of a strongly coupled gauge theory for any length of time. We do not have a complete answer to this question, but it is tempting to speculate that the short-time behavior of the stress-energy tensor is dominated by the singular behavior of the two-point correlation function of components of the stress-energy tensor. In a conformal theory, the correlation functions have a power-law dependence on the invariant space-time distance and diverge on the light cone. This suggests that free streaming of massless ``particles'' may reproduce the short-time behavior of these correlation functions \cite{Calabrese:2005in}.

In summary, we have investigated the evolution of the local pressure anisotropy in a strongly coupled field theory beginning from a spatially inhomogeneous, far off-equilibrium initial state. The anisotropy can be well described at early times by a free-streaming model. Towards the end of the early time-window we can study, the stress-tensor is consistent with the second order hydrodynamic stress tensor computed from the local energy density and fluid velocity.
Our results are complementary to those of \cite{Heller:2011ju}, who investigated the approach to hydrodynamics for a spatially homogeneous, but longitudinally expanding fluid. Taken together, these two results suggest that free streaming followed by second order viscous hydrodynamics may provide a valid description of the space-time evolution of the stress energy tensor in strongly coupled conformal gauge theories over the entire history. This would justify the standard approach to early time evolution of partonic matter in relativistic heavy ion collisions. However, given the very short time window our methods allow us to describe, more work is required to confirm whether the early-time agreement with second order hydrodynamics that we find is accidental or really signals the onset of a hydrodynamic regime.
Specifically, we cannot exclude the possible onset of violent equilibration dynamics at later times, as found in \cite{vanderSchee:2013pia} for central, homogeneous, boost-invariant collisions. It would also be interesting to investigate different classes of initial conditions which prepare the non-equilibrium state in qualitatively different ways (e.g., IR injection of energy).  It is possible that some of these classes would lead to different conclusions regarding, e.g., the validity of free streaming at early times.

{\em Acknowledgments:} We thank M.~Heller, K.~Rajagopal and especially S.~Minwalla for discussions. This research is supported  by the Belgian Federal Science Policy Office, by FWO-Vlaanderen, by the Foundation of Fundamental Research on Matter (FOM), by  Finnish Academy of Science and Letters, by the U.S. Department of Energy, by the BMBF, and by Academy of Finland. AB is a postdoctoral researcher FWO and FG is Aspirant FWO.


\end{document}